\documentclass[twocolumn,aps,prc,superscriptaddress,nofootinbib,showpacs,floatfix,singlecolumn]{revtex4}
\usepackage{url}
\usepackage{cancel}
\usepackage[colorlinks,linkcolor=blue,citecolor=blue,filecolor=black,urlcolor=blue]{hyperref}
\usepackage{epsfig,graphics}
\usepackage{graphicx}
\usepackage{dcolumn}
\usepackage{bm}
\usepackage[usenames]{color}
\usepackage{amssymb}
\usepackage{amsmath}
\usepackage{multirow}
\usepackage{float}
\usepackage{harpoon}
\usepackage{MnSymbol}
\usepackage{appendix}
\usepackage{color}

\makeatletter

\newcommand{\Rmnum}[1]{\expandafter\@slowromancap\romannumeral #1@}
\makeatother

\makeatletter 
\@addtoreset{equation}{section}
\makeatother  

\newcommand{\sqrtsnn}{\mbox{$\sqrt{s_{\mathrm{NN}}}$}}

\newcommand{\pT} {p_{\mathrm{T}}}
\newcommand{\pPb}{\mbox{$p$+Pb}}
\newcommand{\lr}[1]{\left\langle #1\right\rangle}
\newcommand{\llrr}[1]{\left\llangle #1\right\rrangle}
\newcommand{\Dphi}{\mbox{$\Delta \phi$}}
\newcommand{\Deta}{\mbox{$\Delta \eta$}}
\newcommand{\nch}{N_{\mathrm{ch}}}

\usepackage{CJK}
\hfuzz=\maxdimen
\begin{document}
\title{Influence of initial-state momentum anisotropy on the final-state collectivity in small collision systems}
\newcommand{\sbu}{Department of Chemistry, Stony Brook University, Stony Brook, NY 11794, USA}
\newcommand{\fdu}{Key Laboratory of Nuclear Physics and Ion-beam Application, Institute of Modern Physics, Fudan University, Shanghai 200433, China}
\newcommand{\sdu}{Institute of Frontier and Interdisciplinary Science, Shandong University, Qingdao, 266237, China}
\newcommand{\bnl}{Physics Department, Brookhaven National Laboratory, Upton, NY 11976, USA}
\newcommand{\ccnu}{Institute of Particle Physics, Central China Normal University, Wuhan 430079, China}
\newcommand{\sinap}{Shanghai Institute of Applied Physics, Chinese Academy of Sciences, Shanghai 201800, China}
 \author{Maowu Nie}\affiliation{\sdu}
 \author{Li Yi}\affiliation{\sdu}
 \author{Jiangyong Jia}\email[]{jiangyong.jia@stonybrook.edu}\affiliation{\sbu}\affiliation{\bnl}
 \author{Guoliang Ma}\email[]{glma@fudan.edu.cn}\affiliation{\fdu}\affiliation{\sinap}
  \date{\today} 
\begin{abstract}
A multi-phase transport model is used to understand the origin of long-range collective azimuthal correlations in small-system collisions. To disentangle between collectivity associated with initial-state intrinsic momentum anisotropy and the collectivity arising as a final-state response to the collision geometry, we studied the development of collectivity in 5.02 TeV $p$+Pb collisions with both initial-state and final-state effects included. We find that the initial momentum anisotropy may not be fully isotropized through parton interactions, and the final-state partonic collectivity in general are correlated with both the initial momentum anisotropy and the shape of the collision geometry. The initial momentum anisotropy also influences the event by event fluctuation of collective flow. Therefore the mere evidence of geometry response of the collective flow can not rule out the presence of large contributions from the initial state.
\end{abstract}

\pacs{25.75.Gz, 25.75.Ld, 25.75.-q}
\maketitle
In high-energy hadronic collisions, particle correlations are important tools to study the multi-parton dynamics of QCD in the strongly-coupled non-perturbative regime~\cite{Shuryak:2014zxa}. Measurements of azimuthal correlations have revealed a strong harmonic modulation of particle densities d$N/{\textrm d}\phi\propto 1+2\sum_{n=1}^{\infty}v_{n}\cos n(\phi-\Psi_{n})$, where $v_n$ and $\Psi_n$ represent the magnitude and the phase of the $n^{\mathrm{th}}$-order harmonic, and are often denoted by flow vector $V_n=v_n{\mathrm e}^{{\textrm i}n\Psi_n}$~\cite{Heinz:2013th}. The azimuthal correlations are found to be collective, involving many particles spread over a wide pseudorapidity range. Such azimuthal anisotropy was first observed in large A+A collision system~\cite{Voloshin:2008dg,Heinz:2013th,Luzum:2013yya,Jia:2014jca}, but are then observed and studied in small collision systems such as $pp$ and $\pPb$ collisions at the LHC~\cite{Khachatryan:2010gv,CMS:2012qk,Abelev:2012ola,Aad:2012gla,Aad:2014lta,Khachatryan:2015waa,Aaboud:2017blb,Aaboud:2018syf} and $p$+Au, $d$+Au and $^3$He+Au collisions at RHIC~\cite{Adare:2013piz,Adare:2014keg,Adamczyk:2015xjc,Aidala:2017ajz,PHENIX:2018lia,Huang:2019rsn}. 

Although azimuthal anisotropy in A+A collisions is naturally explained as a result of hydrodynamic collective expansion of the hot and dense matter produced in the collision~\cite{Dusling:2015gta}, the applicability of hydrodynamic picture for the azimuthal anisotropy in small collision systems such as $pp$ or $p$+A collisions remains an open question~\cite{Nagle:2018ybc,Antinori:2019hag,Kurkela:2019kip}. It has been argued that the size is too small and life-time is too short for the matter in small system to hydrodynamize and approach local isotropization~\cite{Schenke:2017bog}. Instead, the azimuthal anisotropy may reflect intrinsic long-range momentum correlations of the dense gluon field right after the collision~\cite{Dusling:2013qoz,Gyulassy:2014cfa,Blok:2017pui}. The current debate is focused on the timescale for the emergence of collectivity: Is the collectivity born in the initial state, developed during non-equilibrium transport before the system hydrodynamizes, or arises even later when the system can be described by hydrodynamics? The latter two scenarios lead to a collectivity that correlates with the initial spatial eccentricities, while the first does not. 

The system produced right after the collision is highly anisotropic in momentum space, and the strong interactions among the constituents of the produce system tend to isotropize this anisotropy~\cite{Strickland:2014pga,Romatschke:2017ejr}. However, the short life-time may prevent the produce matter to fully isotropize before hadronization. In this case, the collectivity of final-state particles may has contributions from both initial momentum anisotropy and final-state geometry-driven anisotropy~\cite{Zhang:2015cya,Greif:2017bnr,Kurkela:2019kip}. In this Letter, we investigate the possibility and consequence of coexistence of initial-state and final-state effects using a transport model for $p$+Pb collisions at $\sqrtsnn=5.02$ TeV. We show that the long-range azimuthal anisotropy for final-state particles could be strongly modified by the initial momentum anisotropy while it still maintains a strong correlation with the initial spatial eccentricity. We find that the short-range azimuthal anisotropy is sensitive to microscopic mechanism for initial momentum anisotropy or final-state non-equilibrium effects. 

The model used for this study is a multi-phase transport model (AMPT)~\cite{Lin:2004en}. The AMPT model is successful in describing several features of small-system collectivity at RHIC and the LHC, over a wide range of collision species and energies~\cite{Adare:2015cpn,Ma:2014pva,Bzdak:2014dia,Nie:2018xog}. It starts with Monte Carlo Glauber initial conditions, the space-time evolution of the collision is modeled via strings and jets that melt into partons, followed by parton scattering, parton coalescence, and hadronic scattering. The collectivity is generated mainly in the partonic scattering stage, known as the Zhang's Parton Cascade (ZPC), which leads to an azimuthal anisotropy of final particles correlated with the shape of the initial geometry. We use the setup of Ref.~\cite{Nie:2018xog} with a partonic cross-section of 3~mb.

In this Letter, we focus on the leading component of azimuthal anisotropy, elliptic flow $V_2= v_2 e^{{\textrm i}2\Psi_2}$. Since the number of particles in each event is finite, the $V_2$ can only be estimated from the $\phi$ angle of the particles,
\begin{equation}\label{eq:1}
{\bm q}_2 \equiv q_2 e^{i2\Psi_2^{\mathrm{obs}}} =  \lr{e^{i2\phi}}\;,
\end{equation}
where the observed event plane $\Psi_2^{\mathrm{obs}}$ smears around the true EP angle $\Psi_2$ due to statistical fluctuations.

In the final-state scenarios, $V_2$ is driven by the eccentricity vector ${\mathcal{E}}_2$, which can be calculated from initial-state coordinates $(r_i,\phi_i)$ of the participant nucleons
\begin{equation}\label{eq:2}
{\mathcal{E}}_2 \equiv\varepsilon_2 {\mathrm e}^{{\textrm i}2\Psi^{\mathrm{PP}}_2} = -\frac{\lr{r^2e^{i2\phi}}}{\lr{r^2}}, 
\end{equation}
where the $\Psi^{\mathrm{PP}}_2$ is known as the participant plane (PP). 

To precisely control the amount of initial momentum anisotropy, a two-step procedure is used to prepare partons entering the ZPC stage. We first randomize the $\phi$ angle of all initial partons to eliminate any global initial momentum anisotropy (but the momentum distribution in local rest frame is generally anisotropic). This step also removes any azimuthal anisotropy associated with non-flow effects. We then rotate the parton $\phi$ angle via the procedure described in Ref.~\cite{Masera:2009zz} to produce a fixed amount of initial elliptic anisotropy along a random event-wise direction \mbox{$V_2^{\mathrm{ini}}= v_2^{\mathrm{ini}}e^{i2\Psi_2^{\mathrm{MP}}}$} (MP stands for momentum-plane). These partons then go through the ZPC and later stages of the AMPT. 

In this analysis, we focus on understanding the evolution of collectivity driven by the parton scattering processes. To this end, we calculate and compare the $v_2$ before and after the ZPC using partons in $0.3<\pT<3$ GeV for $p$+Pb collisions at $\sqrtsnn=5.02$ TeV. However, in order to relate to experimental measurements, we present the final results as a function of $\nch$, the number of charged particles in $\pT>0.4$ GeV and $|\eta|<2.5$, after final hadronic transport. 

The elliptic flow coefficient is calculated with both two- and four-particle correlation methods. In the two-particle correlation method, we calculate 
\begin{equation}\label{eq:3}
c_2\{2\}=\llrr{{\mathrm{e}}^{{\textrm i}2(\phi_1^a-\phi_2^b)}}=\lr{v_2^2}\;
\end{equation}
where partons $a$ and $b$ are chosen from two subevents according to $-2.5<\eta_a<-\frac{2.5}{3}$ and $\frac{2.5}{3}<\eta_b<2.5$, and the $\llrr{}$ represent averaging over all pairs in one event then over all events. The large gap between the two subevents reduces the short-range correlations. The flow coefficient from two-particle cumulant, $v_2\{2\}\equiv\sqrt{c_2\{2\}}=\sqrt{\lr{v_2^2}}$, measures the root-mean-square values of $v_2$. Similarly we also calculate $v_2$ using four-particle correlations with the same two subevents~\cite{Jia:2017hbm}:
\begin{align}\nonumber
c_2\{4\} &=\llrr{{\mathrm{e}}^{{\textrm i}2(\phi_1^a+\phi_2^a-\phi_3^b-\phi_4^b)}}-2\llrr{{\mathrm{e}}^{{\textrm i}2(\phi_1^a-\phi_2^b)}}^2\\\label{eq:4}
&=\lr{v_2^4}-2\lr{v_2^2}^2
\end{align}
From this we define the four-particle elliptic flow coefficient $v_2\{4\}$ as
\begin{align}\label{eq:5}
v_2\{4\}=\left(-c_2\{4\}\right)^{1/4}=\left(2\lr{v_2^2}^2-\lr{v_2^4}\right)^{1/4}
\end{align}
which is sensitive to event-by-event fluctuation of elliptic flow. Following Refs.~\cite{Giacalone:2017uqx,Zhou:2018fxx}, we characterize the relative strength of flow fluctuation using a cumulant ratio, 
\begin{align}\label{eq:5b}
nc_2\{4\}\equiv c_2\{4\}/c_2\{2\}^2 = -(v_2\{4\}/v_2\{2\})^4
\end{align}

To further quantify the influence of initial momentum anisotropy and geometry response to the final elliptic flow signal, we also perform a detailed study of the angular correlation between the MP or the PP and the EP of the final-state elliptic flow. The correlation between PP and the EP is calculated as
\small\begin{align}\nonumber
\lr{\cos 2(\Psi_2^{\mathrm{PP}}-\Psi_2)}= \sqrt{\frac{\lr{\cos 2(\Psi_2^{\mathrm{PP}}-\Psi_2^{\mathrm{obs,a}})} \lr{\cos 2(\Psi_2^{\mathrm{PP}}-\Psi_2^{\mathrm{obs,b}})}} {\lr{\cos 2(\Psi_2^{\mathrm{obs,a}}-\Psi_2^{\mathrm{obs,b}})}}}\;.\\\label{eq:6}
\end{align}\normalsize
where $\Psi_2^{\mathrm{obs,a}}$ and $\Psi_2^{\mathrm{obs,b}}$ are the observed EP in subevent $a$ and $b$, respectively. The denominator represents a resolution factor, obtained from the correlation between the two subevents, which corrects for the smearing  of observed EP from the true EP.

Similarly the correlation between the MP for the initial partons and the EP for the final partons is calculated as:
\small\begin{align}\nonumber
\lr{\cos 2(\Psi_2^{\mathrm{MP}}-\Psi_2)} = \sqrt{\frac{\lr{\cos 2(\Psi_2^{\mathrm{MP}}-\Psi_2^{\mathrm{obs,a}})} \lr{\cos 2(\Psi_2^{\mathrm{MP}}-\Psi_2^{\mathrm{obs,b}})}} {\lr{\cos 2(\Psi_2^{\mathrm{obs,a}}-\Psi_2^{\mathrm{obs,b}})}}}.\\\label{eq:7}
\end{align}\normalsize

\begin{figure}[h!]
\begin{center}
\includegraphics[width=1\linewidth]{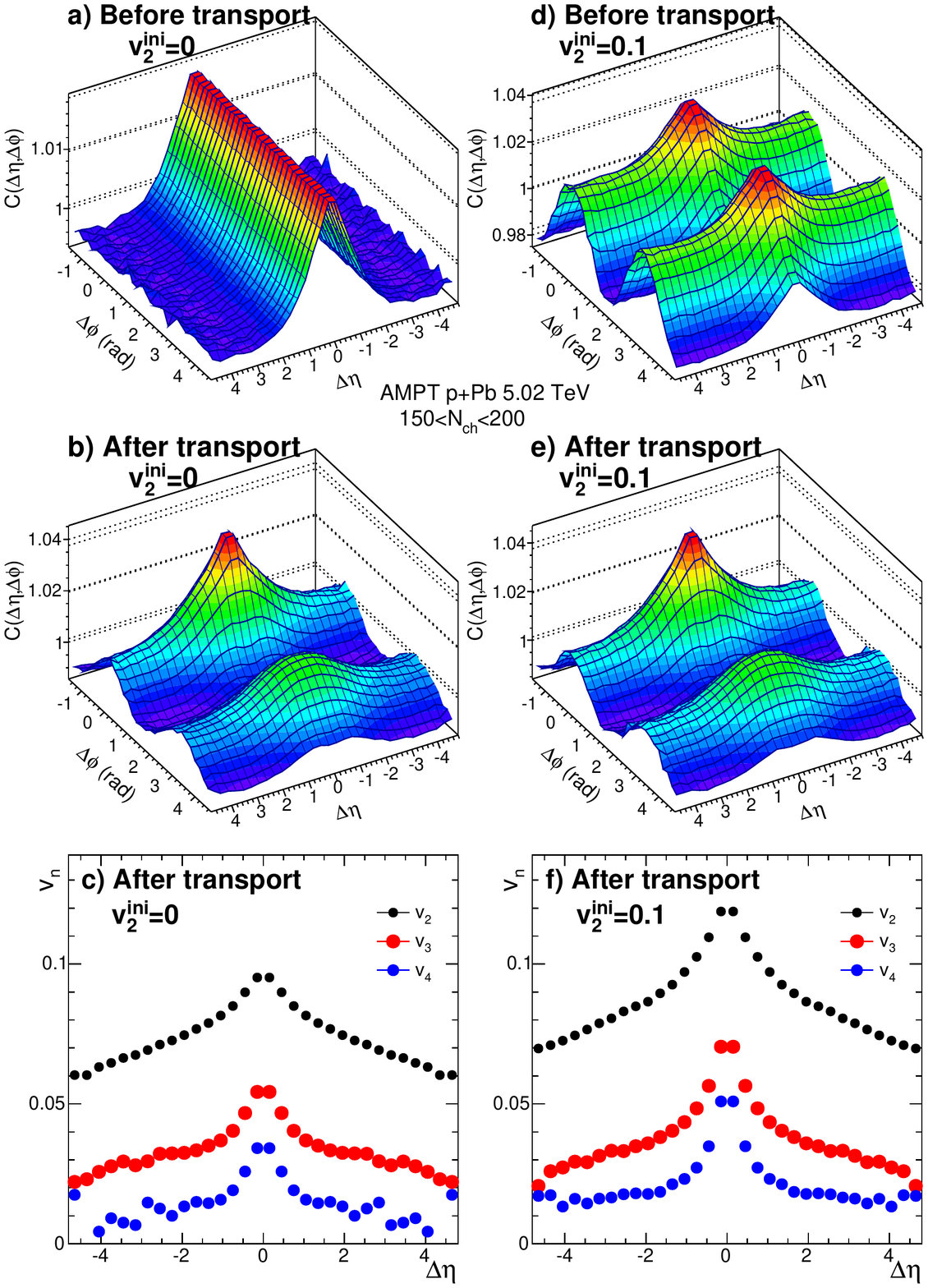}
\end{center}
\vspace*{-0.5cm}
\caption{\label{fig:1} The two-particle correlation in $\Dphi$ and $\Deta$ for initial partons with $v_2^{\mathrm{ini}}=0$ (top-left) and $v_2^{\mathrm{ini}}=0.1$ (top-right) and for final partons after transport with  $v_2^{\mathrm{ini}}=0$ (middle-left) and $v_2^{\mathrm{ini}}=0.1$ (middle-right). The bottom panels shows the $v_n(\Deta)$ for n=2--4 calculated from the corresponding correlation functions in the middle panels.}
\end{figure}

The top and middle rows of Fig.~\ref{fig:1} show the two-particle correlation function for partons in relative azimuthal angle $\Dphi$ and pseudorapidity $\Deta$ before (top row) and after the ZPC (middle row). The correlation function is constructed in the highest multiplicity $p$+Pb collisions ($\nch>150$) as the ratio between pair distribution from the same event and pair distribution from mixed events~\cite{Aad:2014lta}. When the initial anisotropy is set to zero $v_2^{\mathrm{ini}}=0$ (Fig.~\ref{fig:1}(a) ), the distribution is uniform in $\Dphi$. The peak around $\Deta=0$ reflects the short-range correlation which is randomized in $\phi$ but is preserved in $\eta$. After the ZPC (Fig.~\ref{fig:1}(b)), an azimuthal anisotropy develops that appears as a double-ridge at $\Dphi\sim0$ and $\Dphi\sim\pi$ extending to large $\Deta$. On top of the double-ridge are two short-range peaks with different amplitudes at the near and away side.  It is interesting to point out that without final-state interactions, such short-range azimuthal anisotropy, usually associated with ``non-flow'', would not show up in Fig.~\ref{fig:1} (b).

The azimuthal structures of the correlation function is quantified by the Fourier coefficients, $v_n\{2\}^2=\lr{\cos n\Dphi}$, as a function of $\Deta$ in Fig.~\ref{fig:1}(c). The short-range structure in Fig.~\ref{fig:1}(b) is reflected by the narrow peak of $v_n\{2\}$ around $\Deta=0$, on top of a broad distribution associated with the double-ridge in the correlation function. The short-range component in $v_n\{2\}$ appears only after parton scatterings. The broad component in $v_n\{2\}$  shows a slow decrease with $\Deta$, which can be attributed to the longitudinal decorrelation effects~\cite{Bozek:2010vz,Aaboud:2017tql}. 

In a transport picture, two partons with a large $\eta$ separation are causally disconnected and should interact independently with the medium. Therefore, the azimuthal correlation between these two partons arises only through a response to a common geometry shape, leading to the long-range double ridge. For long-range correlations, this mechanism is indistinguishable from a geometry response driven by hydrodynamics. In contrast, the correlation of two partons close to each other in $\eta$ is directly sensitive to the non-equilibrium microscopic scattering processes. The width of the short-range peak reflects the diffusion of the parton in rapidity via random scattering, in a way similar to hydrodynamic fluctuations~\cite{Sakai:2019fpz}. The amplitude of the short-range peak reflects the amount of residual correlation not isotropized by the scattering process. Therefore, the long-range and the short-range correlations together can better constrain hydrodynamics or non-equilibrium transport: the long-range correlation constrains the overall strength of the geometry response, while the short-range correlation is sensitive to the non-equilibrium dynamics.

The right column of Fig.~\ref{fig:1} show the case for $v_2^{\mathrm{ini}}=0.1$. Despite the large initial momentum anisotropy, it only has a modest impact on the correlation function. The corresponding $v_n$ values increase slightly at large $|\Deta|$, while they increase more strongly at $\Deta\sim0$, such that the short-range peak is more evident.

\begin{figure}[h!]
\begin{center}
\includegraphics[width=1\linewidth]{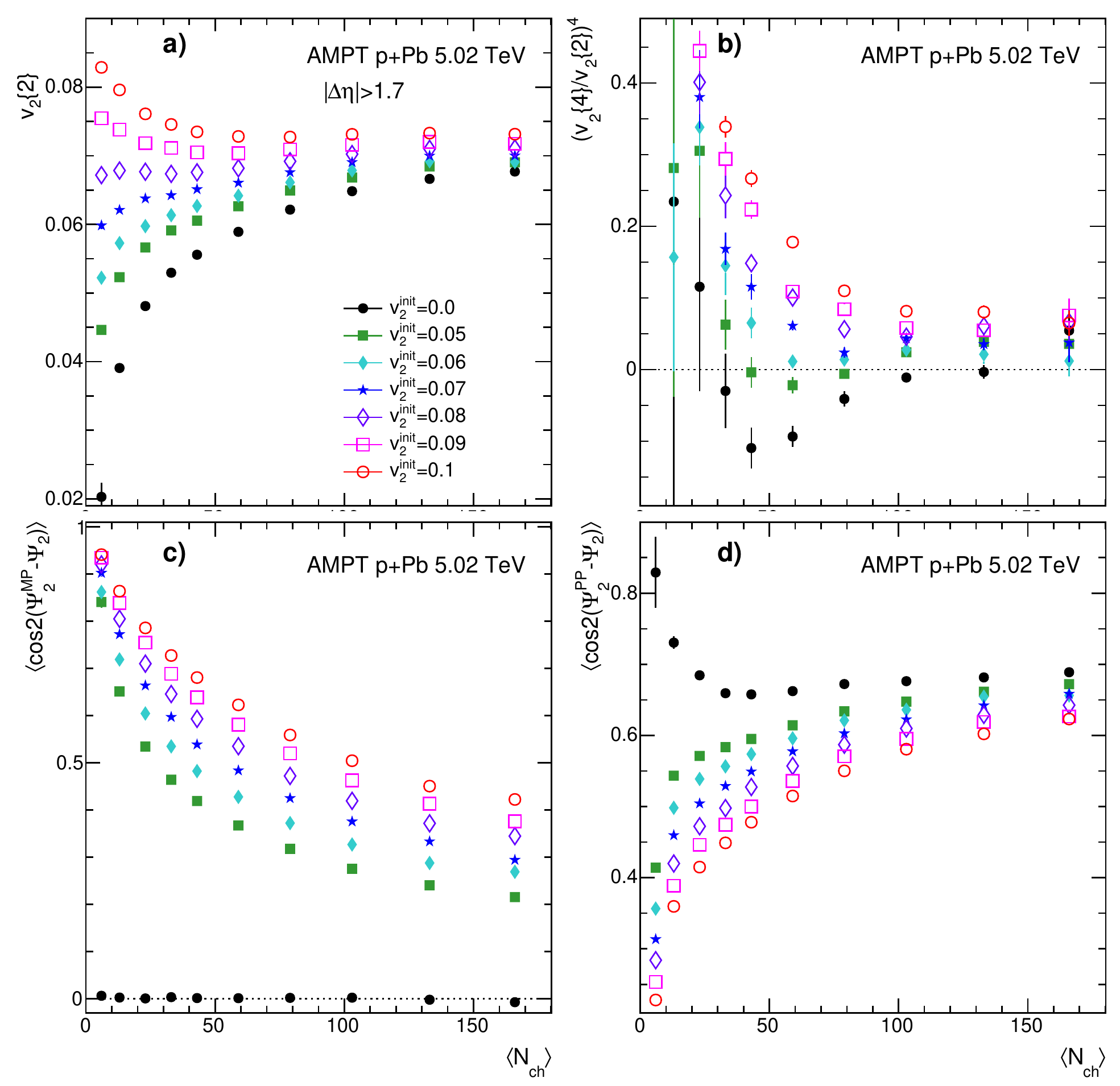}
\end{center}
\vspace*{-0.5cm}
\caption{\label{fig:2} The $\nch$ dependence of final state flow $v_2\{2\}$ (top-left), the relative event by event flow fluctuation via cumulant ratio $(v_2\{4\}/v_2\{2\})^4$ (top-right), angular correlation between the phase of final parton flow and the phase of initial momentum anisotropy (bottom-left), angular correlation between the phase of final parton flow and the PP (bottom-right) for different input values of initial flow.}
\end{figure}

To quantify the final azimuthal anisotropy after ZPC, we calculate the $v_2\{2\}$ and $v_2\{4\}$, and study them as a function of $\nch$. To minimize multiplicity fluctuations, the results are calculated for events with the same $\nch$ and then averaged to obtained results in a finite range of $\nch$. Figure~\ref{fig:2}(a) shows the $\nch$ dependence of $v_2\{2\}$. When $v_2^{\mathrm{ini}}=0$, the $v_2\{2\}$ values show a monotonic increase with $\nch$, reflecting a dominant contribution from eccentricity-driven collective flow. However, as $v_2^{\mathrm{ini}}$ is increased, the $v_2\{2\}$ also increases. The increase is strongest at low $\nch$ region, and is weaker at large $\nch$ region. This behavior implies that at the low $\nch$ region, initial momentum anisotropy dominates the collectivity after the ZPC. At the large $\nch$ region, the initial momentum anisotropy still has up to a 10\% contribution to the final $v_2\{2\}$. 

Figure~\ref{fig:2}(c) shows the correlation between the phase for initial-state partons and the phase for final-state partons: $\cos 2(\Psi_2^{\mathrm{MP}}-\Psi_2)$ (Eq.~\ref{eq:7}). At the low $\nch$ region, the correlator is close to unity, suggesting that the initial momentum anisotropy can easily survive and dominate the final-state elliptic flow. At the high $\nch$ region, the correlation decreases but is still quite large for the $v_2^{\mathrm{ini}}$ values considered. This implies that the final parton's $\Psi_2$ could be strongly biased by the initial momentum anisotropy.

Figure~\ref{fig:2}(d) shows the correlation between the phase of the initial eccentricity and phase of the elliptic flow of final-state partons: $\lr{\cos 2(\Psi_2^{\mathrm{PP}}-\Psi_2)}$ (Eq.~\ref{eq:6}). When $v_2^{\mathrm{ini}}=0$, the correlation decreases with $\nch$, which can be attributed to the fact that $\varepsilon_2$ values also decrease strongly with $\nch$~\cite{Bozek:2011if}. It is more interesting to focus on the trend of the correlator when $v_2^{\mathrm{ini}}$ value is increased. For large $v_2^{\mathrm{ini}}$, the correlator value is dramatically decreased in the low $\nch$ region, and the decrease is smaller at larger $\nch$. This suggests that the $\Psi_2$ is more influenced by the initial momentum anisotropy in the low $\nch$ region, and but is less influenced in the high $\nch$ region.

Finally, Fig.~\ref{fig:2}(b) shows $(v_2\{4\}/v_2\{2\})^4$, the cumulant ratio calculated via Eq.~\ref{eq:5b}. The negative $(v_2\{4\}/v_2\{2\})^4$ values simply imply that $v_2\{4\}$ values become imaginary. The sign and $\nch$-dependent trend of this observable is found to depend on the $v_2^{\mathrm{ini}}$. In particular, large $v_2^{\mathrm{ini}}$ could lead to a negative $c_2\{4\}$ and therefore real $v_2\{4\}$ value. Results in Fig.~\ref{fig:2}(b) imply that the nature of the event-by-event flow fluctuation can be strongly modified in the presence of initial momentum anisotropy.

\begin{figure}[h!]
\begin{center}
\includegraphics[width=1\linewidth]{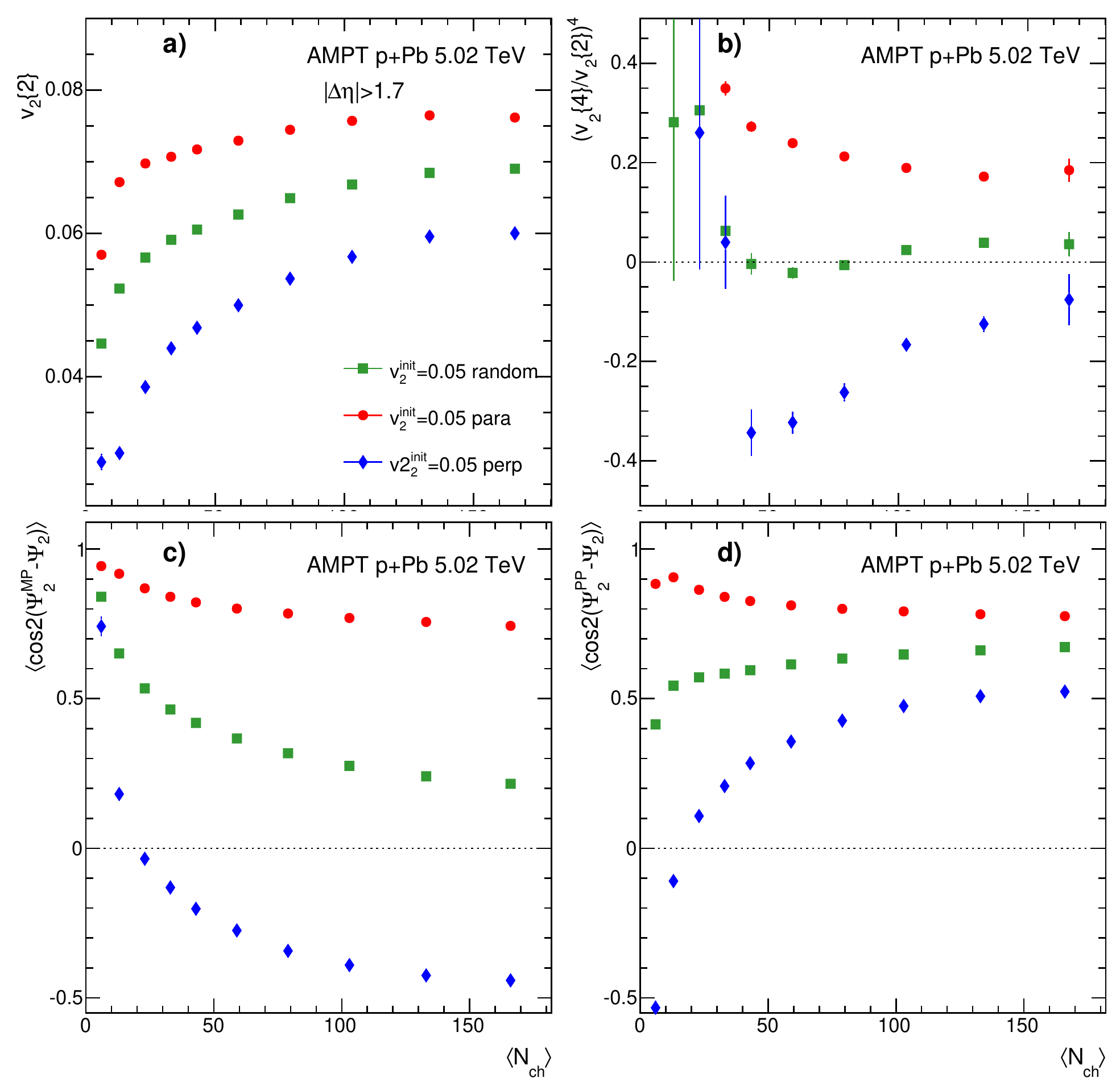}
\end{center}
\vspace*{-0.5cm}
\caption{\label{fig:3} The $\nch$ dependence of final state flow $v_2\{2\}$ (top-left),  the relative event by event flow fluctuation via cumulant ratio $(v_2\{4\}/v_2\{2\})^4$ (top-right), angular correlation between the phase of final parton flow and the phase of initial momentum anisotropy (bottom-left), angular correlation between the phase of final parton flow and the PP (bottom-right) for input initial flow $v_2^{\mathrm{ini}}=0.05$ with three different ways of generating the phase $\Psi_2^{\mathrm{MP}}$: random, parallel $\Psi_2^{\mathrm{MP}}=\Psi_2^{\mathrm{PP}}$ and perpendicular $\Psi_2^{\mathrm{MP}}=\Psi_2^{\mathrm{PP}}+\pi/2$.}
\end{figure}

To further investigate the interplay between the initial momentum anisotropy and geometry-driven flow in the final state, two additional tests with $v_2^{\mathrm{ini}}=0.05$ are carried out. In the first test, the phase of the initial flow $\Psi_2^{\mathrm{MP}}$ is generated to align with the PP: $\Psi_2^{\mathrm{MP}}=\Psi_2^{\mathrm{PP}}$, while in the second test the $\Psi_2^{\mathrm{MP}}$ is generated to be perpendicular to the PP: $\Psi_2^{\mathrm{MP}}=\Psi_2^{\mathrm{PP}}+\pi/2$. The results from these tests are shown in Fig.~\ref{fig:3}. When $\Psi_2^{\mathrm{MP}}$ is aligned with the $\Psi_2^{\mathrm{PP}}$, final-state elliptic flow has larger values (panel a) and its correlations with both $\Psi_2^{\mathrm{MP}}$ and $\Psi_2^{\mathrm{PP}}$ are stronger (panel c and d), and the $(v_2\{4\}/v_2\{2\})^4$ values are positive due to coherent enhancement of elliptic flow from $v_2^{\mathrm{ini}}$ (panel b). When $\Psi_2^{\mathrm{MP}}$ is perpendicular to $\Psi_2^{\mathrm{PP}}$, opposite trends are observed: final-state elliptic flow has smaller values and its correlations with both $\Psi_2^{\mathrm{MP}}$ and $\Psi_2^{\mathrm{PP}}$ are weakened, and the $(v_2\{4\}/v_2\{2\})^4$ values become negative.

Recently, the possibility for a further scan of small collision systems at RHIC and LHC has been discussed~\cite{Rybczynski:2017nrx,Citron:2018lsq,Sievert:2019zjr,Lim:2018huo,Huang:2019tgz}. The system scan at fixed $\sqrtsnn$ varies mainly the size and shape of the initial fireball, while RHIC-LHC energy scan for a fixed system such as O+O provides a setup with the same nucleon geometry but much larger parton densities at LHC. Therefore such system and energy scan allow us to vary the role of initial momentum anisotropy, non-equilibrium transport and hydrodynamics, and then study the change in $v_n$. For example, the ordering of $v_n$ at fixed $\nch$ between different systems or different energies can provide additional sensitivity on the initial momentum anisotropy, if such ordering do not follow the expected scaling from initial eccentricities (see Fig.~\ref{fig:2}(a)). Furthermore, since the long-range correlation constrains the overall strength of the geometry response and the short-range correlation constrains the non-equilibrium dynamics (see Fig~\ref{fig:1}), a simultaneous study of the long-range and short-range correlations can be used to disentangle two competing geometry response models based on non-equilibrium transport or hydrodynamics. 

In summary, we studied the influence of the initial momentum anisotropy to the geometry-driven collective flow in $\pPb$ collisions at 5.02~TeV using the AMPT transport model. We find that the initial momentum anisotropy may not be fully isotropized during partonic transport, and the final collective flow is correlated with directions of both the initial momentum anisotropy and the geometrical eccentricity. The presence of initial momentum anisotropy also changes dramatically the strength of the event-by-event flow fluctuations. Therefore, the mere evidence of geometry response of the collective flow can not rule out presence of large contribution from the initial state. A more comprehensive small system scan mapping out detailed pseudorapidity structures of two- and multi-particle correlation observables are necessary to quantify the contributions from different scenarios. 

We appreciate valuable discussion with Roy Lacey and Zhenyu Chen.  This work is supported by NSF grant number PHY-1613294 (JJ),  NSFC grant number 11890714, 11835002, 11421505, and the Key Research Program of the Chinese Academy of Sciences grant Number XDPB09 (G.L.M), NSFC grant number 11890713 (MN,YL).

\bibliography{ref}{}
\bibliographystyle{apsrev4-1}
\end{document}